\documentclass[12pt]{article}
\usepackage{paper2e}
\usepackage{mydefs2e}

\makeatletter
        \@addtoreset{equation}{section}%
\makeatother

\renewcommand{\Yasymm}{\raisebox{-3pt}{\drawsquare{7}{0.6}\hskip-7.6pt%
\raisebox{7pt}{\drawsquare{7}{0.6}}}}

\newcommand{\thooft}{'t~Hooft\xspace}

\begin{document}
\begin{titlepage}
\preprint{UMD-PP 97-132\\
hep-ph/9706554}

\title{Simple Gauge-mediated Models\\\medskip
with Local Minima}

\author{Markus A.~Luty}

\address{Department of Physics, University of Maryland\\
College Park, Maryland 20742, USA\\
{\tt mluty@physics.umd.edu}}

\begin{abstract}
We describe a simple class of \susc gauge theories that can act as
\susy-breaking sectors for gauge-mediated \susy breaking.
The models have a local \susy-breaking minimum along a direction
in field space where a singlet gets a large expectation value.
The potential along this direction has a runaway behavior
stabilized by \susy breaking in the effective low-energy theory.
The \susc vacua are at infinite field values, and
cosmological bounds on false vacuum decay are easily satisfied.
The models have no dimensionful parameters, and all mass scales arise
through strong coupling dynamics.
Simple variants of the model are compatible with perturbative
unification, can naturally have dynamical \susy breaking
at a scale as low as $10\TeV$, and can solve the $R$ axion problem
without appealing to Planck-scale effects.
\end{abstract}

\date{June, 1997}

\end{titlepage}

\noindent
Gauge-mediated \susy breaking is an interesting alternative to
traditional supergravity-mediated \susy breaking scenarios that
has received renewed attention recently
\cite{OldGaugeMediated,NewGaugeMediated}.
Many explicit models of gauge-mediation have appeared in the
literature \cite{NewGaugeMediated}, but early models were very
complicated and may best be regarded as existence proofs.
(For recent progress in model-building, see
\Refs{PTtype,Randall,hitoshi,DDRG}.)
In this paper, we construct a simple class of gauge-mediated models
that naturally generate a \emph{local} \susy-breaking minimum for
a singlet.
These models have a classical flat direction along which a singlet
$S$ gets a large \vev, and many fields get massive.
The flat direction is lifted by two competing quantum effects:
gaugino condensation in one group factor generates a runaway
dynamical superpotential for $S$, while dynamical supersymmetry
breaking in another group factor stabilizes the potential.
These models have many similarities with the models of
Murayama \cite{hitoshi} and
Dimopoulos, Dvali, Rattazzi, and Giudice \cite{DDRG},
which utilize a flat \susy-breaking potential generated by strong
dynamics in one gauge group factor, stabilized by perturbative dynamics
in another gauge group factor.

To illustrate the idea of the present class of models, consider a theory
with symmetry group
\beq
G = \SU{N}_P \times \SU{5}_B \times [\SU{K} \times \SU{K'}],
\eeq
where the group factors on the left are gauge symmetries, and the
group in brackets is a global symmetry.
(We will eventually gauge part of the global symmetry with the
standard-model gauge group.)
The matter content is
\beq\bal
P &\sim (\Yfund, 1) \times (\Yfund, 1),
\\
\bar{P} &\sim (\bar{\Yfund}, 1) \times (\bar{\Yfund}, 1),
\\
Q &\sim (1, \Yfund) \times (1, \Yfund),
\\
\bar{Q} &\sim (1, \bar{\Yfund}) \times (1, \bar{\Yfund}),
\\
\bar{T} &\sim (1, \bar{\Yfund}) \times (1, 1),
\\
A &\sim (1, \Yasymm) \times (1, 1),
\\
S &\sim (1, 1) \times (1, 1).
\eal\eeq
The theory has a superpotential
\beq
W = \la S \tr(P \bar{P})
+ \ka S \tr(Q \bar{Q}),
\eeq
where the trace is over the flavor indices.
This is the most general renormalizable superpotential consistent
with the symmetries given above together with a $\U1_R$ symmetry under
which $S$ has charge $+2$, so this theory is strongly natural in the
sense of \thooft.
As we will see, the role of the $\SU{N}_P$ gauge group is to push $S$
away from the origin, and the role of $\SU{5}_B$ is to break \susy.
The couplings of the two gauge factors may be defined by the scales
$\La_P$ and $\La_B$ (analogous to $\La_{\rm QCD}$) where the
perturbative gauge couplings become strong.
%
%
We will assume that $\SU{N}_P$ is stronger than $\SU{5}_B$, that is,
$\La_P \gg \La_B$.

This theory has a classical flat direction where $S$ is nonzero and
all other fields vanish.
We will show that the theory has a \emph{local} minimum for large
values of $S$ along this direction.
For $S \ne 0$, the fields $P, \bar{P}$ and $Q, \bar{Q}$
become massive.
We will write a low-energy effective theory for the fields that are
massless at the classical level.
These include the $\SU{N}_P \times \SU{5}_B$ gauge multiplets, and the
fields $S$, $\bar{T}$, and $A$ transforming under $\SU{5}_B$.
(When we gauge part of the global symmetry with the standard-model
gauge group, the standard-model fields are of course also light.)
It is important that for large $S$, the $\SU{5}_B$ $D$ terms lift all
flat directions in the fields $A$ and $\bar{T}$, so we have only
a one-parameter flat direction.

Since $\Lambda_N \gg \Lambda_5$, the most important effect in the
low-energy theory is gaugino condensation in the $\SU{N}_P$ group.
The $\SU{N}_P$ dynamics becomes strong at a scale that depends on $S$
through the one-loop matching at the scale where $P, \bar{P}$
become heavy:
\beq[matchone]
\La_{P,{\rm eff}}(S) = \La_P^{1 - K / (3N)} (\la S)^{K / (3N)}.
\eeq
The effective lagrangian is an expansion in inverse powers of $S$,
since it arises from integrating out modes with mass proportional to
$S$.
Gaugino condensation does not break \susy, so this gives rise to
\susc terms in the effective lagrangian
\beq[effone]
\bal
\scr{L}_{\rm eff} &\simeq \frac{1}{16\pi^2} \biggl\{
\myint d^2\th\, \La_{P,{\rm eff}}^3(S) + \hc
\\
&\qquad\qquad
+ \myint d^4\th\, \left( | \La_{P,{\rm eff}}(S) |^2
+ \frac{| \La_{P,{\rm eff}}(S) |^4}{|\la S|^2} + \cdots \right)
\biggr\},
\eal
\eeq
where we have omitted unknown order-1 coefficients, as we
do throughout the paper.
(For an explanation of the factors of $4\pi$ in strongly interacting
\susc theories, see \Refs{fourpi}.)
Using \Eq{matchone}, the dynamical superpotential for $S$ is \cite{ADS}
\beq
W_{\rm dyn} \simeq \frac{1}{16\pi^2} \La_P^3
\left( \frac{\la S}{\La_P} \right)^{K/N},
\eeq
We will assume that $K < N$, 
so that this superpotential forces $S$ to run away.

The $\SU{5}_B$ dynamics becomes strong at a scale
\beq[matchfive]
\La_{B,{\rm eff}}(S) = \La_B^{1 - K'/13} (\ka S)^{K'/13}.
\eeq
The strong $\SU{5}_B$ dynamics is believed to break \susy
\cite{fivebreak}.
This theory has no small parameters, and so the \susy breaking is
parameterized in the effective lagrangian by the terms
\beq[fivebreak]
\de\scr{L}_{\rm eff} \simeq \frac{1}{16\pi^2} \biggl\{
| \La_{B,{\rm eff}}(S) |^4
+ \frac{| \La_{B,{\rm eff}}(S) |^6}{|\ka S|^2}
+ \cdots \biggr\}.
\eeq
(The Goldstino generated by the dynamics is an additional light field
that is not shown explicitly here.
The Goldstino couplings can be included using standard methods
\cite{Gold}.)
From \Eq{matchfive}, we see that the first term in \Eq{fivebreak}
grows with $S$ and stabilizes the runaway potential, so that
\susy is broken at the (local) minimum.

To minimize the potential, we integrate out $F_S$ to obtain
\beq
\bal
V(S) &\simeq \frac{\la^{2K / N} \La_P^4}{(16\pi^2)^2} \frac{K^2}{N^2}
\left( \frac{\La_P}{S} \right)^{2(N - K)/N}
+ \cdots
\\
&\qquad
+ \frac{\ka^{4 K' / 13} \La_B^4}{16\pi^2}
\left( \frac{S}{\La_B} \right)^{4 K' / 13}
+ \cdots.
\eal
\eeq
The omitted terms arise from terms in the effective lagrangian
that are suppressed by additional powers of $S$ compared to those
shown.
(It must be checked that these are in fact negligible when we minimize the
potential.)
Neglecting higher-order terms, we obtain a local \susy-breaking minimum at
\beq[svev]
\avg{S} \simeq \La_P \left[ 
\frac{13 K^2(N - K)}{32 \pi^2 N^3 K'}
\frac{\la^{2 K / N}}{\ka^{4 K' / 13}}
\left( \frac{\La_P}{\La_B} \right)^{4 (13 - K') / 13}
\right]^n,
\eeq
where
\beq
n = \frac{13 N}{2[2 K' N + 13(N - K)]}.
\eeq
$n > 0$ because $K < N$, so $\avg{S} \gg \La_P$ as
long as $\La_P \gg \La_B$.
The $F$ component of $S$ gets a \vev
\beq[fsvev]
\frac{\avg{F_S}}{\avg{S}^2} \simeq
\frac{\la^{K/N}}{16\pi^2} \frac{K}{N} \left[
\frac{32\pi^2 N^3 K'}{13 K^2 (N - K)}
\frac{\ka^{4 K' / 13}}{\la^{2 K / N}}
\left( \frac{\La_B}{\La_P} \right)^{4 (13 - K') / 13}
\right]^{(3 N - K) n / N}.
\eeq
We see that $\avg{F_S}/\avg{S}^2$ is proportional to a
positive power of $\La_B / \La_P$, so
$\avg{F_S} \ll \avg{S}^2$ as long as $\La_B \ll \La_P$.
This fact, together with
$\avg{S} \gg \La_{P,{\rm eff}}(\avg{S}) \gg \La_{B,{\rm eff}}(\avg{S})$
ensures that the terms
neglected in minimizing the potential are in fact negligible.
The complicated powers in \Eqs{svev} and \eq{fsvev} come
from the fact that all mass scales in this model arise
from dimensional transmutation.

The Goldstino arises dominantly from the $\SU{5}_B$ symmetry breaking
sector.
The Goldstino decay constant $F$ is therefore given by
\beq[goldf]
F \simeq \frac{\La_{B,{\rm eff}}^2(\avg{S})}{4\pi}
\simeq \left( \frac{13 (N - K)}{2 N K'} \right)^{1/2} \avg{F_S},
\eeq
and we see that $F \sim \avg{F_S}$.

If we gauge a subgroup of the global symmetry with
the standard-model gauge group, this model corresponds precisely to
a minimal messenger model \cite{NewGaugeMediated}.
The fields $P, \bar{P}$ and/or $Q, \bar{Q}$ carry standard-model quantum
numbers, and play the role of the messengers.
The soft \susy breaking terms in the observable sector are proportional
to the ``messenger scale''
\beq
M_{\rm mess} \equiv \frac{\avg{F_S}}{\avg{S}}.
\eeq
Numerically, we require $M_{\rm mess} \sim 10\TeV$ in order to
obtain realistic superpartner masses.
It can be checked that the masses of the $S$ scalars and vectors are
also of order the messenger scale in this model.
For example, the $S$ fermion mass is
\beq[smass]
m_{\psi_S} \simeq \frac{N - K}{N} M_{\rm mess}.
\eeq
These results show that the only important scales in the
model are $\avg{S}$ and $\avg{F_S}$.
These scales are complicated functions of the parameters of the
model (see \Eqs{svev} and \eq{fsvev}), but all other mass scales are
proportional to simple ratios of these scales.

In order to get some additional intuition about the scales involved,
we consider two special cases.
The first case is the ``minimal scenario'' where
\beq
N = 2,
\quad
K = 1,
\quad
K' = 5.
\eeq
This is the smallest model that has an $\SU{5}$ global symmetry that can
be gauged with the standard-model gauge group.
(The standard-model gauge group $\SU{3}_C \times \SU{2}_W \times \U1_Y$
is imbedded into $\SU{5}$ in the standard way, and we refer to this
imbedding as $\SU{5}_{\rm SM}$ for brevity.)
The model contains $5 \times (\Yfund \oplus \bar{\Yfund})$ under
$\SU{5}_{\rm SM}$, so this model is marginally compatible with perturbative
unification for low values of $\avg{S}$.
In this case,
\beq
\frac{\avg{S}}{\La_P} \simeq 0.07\,
\frac{\la^{0.39}}{\ka^{0.61}}
\left( \frac{\La_P}{\La_B} \right)^{0.97},
\qquad
\frac{\avg{F_S}}{\avg{S}^2} \simeq 3\,
\frac{\ka^{1.5}}{\la^{0.48}}
\left( \frac{\La_B}{\La_P} \right)^{2.4}.
\eeq

The second case is the ``maximal scenario'', where the $\SU{5}_B$
gauge factor has so many matter fields that it is barely
asymptotically free:
\beq
N = 2,
\quad
K = 1,
\quad
K' = 12.
\eeq
In this model, the $\SU{5}_B$ gauge coupling runs very slowly until the
scale $\ka S$, where many $\SU{5}_B$ charged fields become massive;
below this scale, the effective $\SU{5}_B$ coupling becomes strong
quickly.
This mechanism is interesting because the effective $\SU{5}_B$
dynamical scale is tied closely to $\La_P$, and
(to a good approximation) the model has only one scale.
If we gauge an $\SU{5}_{\rm SM}$ subgroup of the global $\SU{12}$ symmetry,
then this model also contains $5 \times (\Yfund \oplus \bar{\Yfund})$ under
$\SU{5}_{\rm SM}$.
In this case,
\beq
\frac{\avg{S}}{\La_P} \simeq 0.2\,
\frac{\la^{0.21}}{\ka^{0.79}}
\left( \frac{\La_P}{\La_B} \right)^{0.066},
\qquad
\frac{\avg{F_S}}{\avg{S}^2} \simeq 0.2\,
\frac{\ka^{2.0}}{\la^{0.033}}
\left( \frac{\La_B}{\La_P} \right)^{0.16}.
\eeq
As expected, the physical results are very insensitive to $\La_B$.
However, our approximations are only valid if
$\la \avg{S} \gg \La_P$ and $\avg{F_S} \ll \avg{S}^2$.
Both of these conditions are satisfied for $\la \sim 1$, $\ka \ll 1$.
We can also consider the regime where $\la \sim \ka \sim 1$,
assuming that the local minimum persists despite the fact that our
approximations are no longer under control.
This is interesting because in this regime the model effectively has a
single scale and no small parameters, and the \susy-breaking dynamics
takes place at energies as low as $\La_P \sim 10\TeV$.
This holds out the hope that dynamical \susy
breaking may be accessible to direct experimental study.

We now consider the global minimum in this model.
The classical space of vacua has two branches.
For $S \ne 0$, we have the one-parameter space of classical flat
directions analyzed above; this branch has no \susc minimum when
quantum effects are included.
We now consider the $S = 0$ branch, taking $K = 1$ for simplicity.
Along this branch, the $\SU{N}_P$ gauge dynamics gives rise to a
dynamical superpotential in the direction parameterized by the
gauge-invariant $X \equiv P \bar{P}$.
The effective superpotential below the scale $X$ is then
\beq
W_{\rm eff} \simeq \la S X + \ka S \tr(Q \bar{Q})
+ \frac{\La_P^3}{16\pi^2} \left(
\frac{\La_P^{2}}{X} \right)^{1/(N - 1)}.
\eeq
The description in terms of $X$ is smooth because we are considering
vacua with $X \ne 0$.
We can therefore integrate out $S$ and $X$ to obtain an effective
$\SU{5}_B$ gauge theory with superpotential
\beq
W_{\rm eff} \simeq \frac{\La_P^3}{16\pi^2} \left(
\frac{\la \La_P^{2}}{\ka \tr(Q \bar{Q})} \right)^{1/(N - 1)}.
\eeq
One can check that all \susc vacua in this model have at least one
runaway direction.%
\footnote{For the dynamics of the theory with $K' \le 4$, see
\Refs{PoppitzTrevedi,Pouliot}.
The dynamics of this theory for $K' \ge 5$ is presently not understood.
(One can write a ``dual'' for this theory using deconfinement
\cite{Pouliot}, but it is not weakly coupled.)
However, symmetry and holomorphy do not allow the moduli space to be
modified at the quantum level for $K' \ge 5$, so it is clear that the
theory has a moduli space of \susc vacuum states with
$\tr(Q\bar{Q}) \to \infty$, even if we do not understand the physics of
many of these vacua.}
This is already an interesting result, since it shows that the
local \susy-breaking minimum is closer to the origin of field space
than the \susc vacua, making it more likely that the universe
ends up in the false vacuum if it cools from temperatures larger
than $\La_P$.
(In an inflationary universe, these considerations apply only
if the reheat temperature is larger than $\La_P$.)

We would like to know the lifetime of the false vacuum.
The potential in this model is far too complicated to do an honest
calculation, so we will limit ourselves to simple estimates.
Because the energy difference between the false vacuum and the
true vacuum is small compared to the distance in field space to the
classical escape point, we can bound the tunnelling rate
by approximating the potential as completely flat.
In that case, the Euclidean tunnelling action is
\cite{flatbounce,DDRG}
\beq
I_{E} \simeq 2\pi^2 \frac{(\De\phi)^4}{V},
\eeq
where $\De\phi$ is the distance in field space to the classical escape
point, and $V$ is the value of the potential in the false vacuum.
The value of $Q$ in the $S = 0$ branch where the energy density is
equal to that of the false vacuum is
\beq
Q \sim \avg{S} \left(
\frac{\La_P}{\avg{S}} \right)^{[2N + K(N - 1)]/[N(N + 1)]} \ll \avg{S}.
\eeq
Therefore, our estimate is dominated by the tunnelling from the false
vacuum to $S \sim \La_P$.
In this way we obtain the bound
\beq
I_E \gsim 2\pi^2 \frac{\avg{S}^4}{\avg{F_S}^2},
\eeq
which is always large.
Since this is an extremely conservative estimate, we believe that
decay into the \susc vacuum is not a problem in these models.

It should be clear that the ideas involved in the model described
above are very robust, and it is easy to construct variants.
Many different gauge theories could act as the ``push''
and \susy-breaking sectors of a model of this type.
We close by mentioning some generalizations that have additional
phenomenologically attractive features.
First, if we want to preserve perturbative one-step unification, we can
introduce a single $\Yfund \oplus \bar{\Yfund}$ of $\SU{5}_{\rm SM}$
coupled to $S$ instead of gauging a subgroup of the global
$\SU{K} \times \SU{K'}$ symmetry.
These messengers get mass for $S \ne 0$, and the analysis of the local
minimum proceeds exactly as above.

Another variant of the above model solves the $R$-axion problem without
appealing to Planck-scale effects \cite{Rax}.
The idea is to introduce several ``push'' group factors with different
numbers of colors and flavors to break the $\U1_R$ symmetry explicitly.
For example, a model with ``push'' group $\SU{N_1} \times \SU{N_2}$ with
$K_1$ and $K_2$ flavors, respectively,
has no anomaly-free $\U1_R$ symmetry as long as
$K_1/N_1 \ne K_2/N_2$.
If $K_1 < N_1$ and $K_2 < N_2$,
the dynamical superpotential along the $S$
classical flat direction is given by gaugino condensation in the
groups $\SU{N_1}$ and $\SU{N_2}$:
\beq
W_{\rm eff} \simeq \frac{1}{16\pi^2} \left[
\La_{P1}^3 \left( \frac{\la_1 S}{\La_{P1}} \right)^{K_1 / N_1}
+ \La_{P2}^3 \left( \frac{\la_2 S}{\La_{P2}} \right)^{K_2 / N_2}
\right].
\eeq
This clearly has no $\U1_R$ symmetry.
In general, there will be a hierarchy between the scales $\La_{P1}$ and
$\La_{P2}$, so there will be an approximate $\U1_R$ symmetry and
the $R$ axion will be a pseudo Nambu--Goldstone mode.
In this model, if we assume $\La_{P2} \ll \La_{P1}$, the $R$ axion
mass is
\beq
m_R \sim M_{\rm mess}
\frac{\la_2^{K_2 / (2 N_1)}}{\la_1^{K_1 / (2 N_2)}}
\left( \frac{\La_{P2}}{\La_{P1}} \right)^{(3 N_2 - K_2) / N_2}
\left( \frac{\avg{S}}{\La_{P1}} \right)^{K_2/N_2 - K_1/N_1}.
\eeq
This can be phenomenologically acceptable for a wide range of parameters.
\Ref{Rax} showed that fine-tuning the cosmological constant in supergravity
breaks $\U1_R$ explicitly, leading to an $R$ axion mass.
In the present model, this mass is
\beq
m_{R,{\rm grav}} \sim \left(
\frac{\avg{F_S}^2}{\avg{S} M_{\rm Planck}} \right)^{1/2}
\sim M_{\rm mess} \left(
\frac{\avg{S}}{M_{\rm Planck}} \right)^{1/2},
\eeq
which is always larger than $100\MeV$ even if
$\avg{S} \sim \avg{F_S}^{1/2} \sim 10\TeV$.
This is safe from supernova constraints, so it is not clear that
the $\U1_R$ breaking mechanism discussed above
is required in these models.

\section*{Acknowledgments}
I thank A. Kusenko, J. Terning, and especially R. Rattazzi for helpful discussions,
and the CERN theory group for hospitality while this work was being
carried out.


\end{document}